\documentclass[12pt]{article}
\usepackage{rotate}

\input epsf

\newenvironment{sciabstract}{%
\begin{quote} \bf}
{\end{quote}}

\newcounter{lastnote}

\title{Using Sequence Alignments to Predict Protein Structure and Stability With High Accuracy }

\author
{Alan Lapedes,$^{1, 2, \ast}$ Bertrand Giraud,$^{3}$ Christopher Jarzynski $^{1}$\\
\\
\normalsize{$^{1}$Theoretical Division, Los Alamos National Laboratory }\\
\normalsize{MS B213, Los Alamos N.M. 87545 USA}\\
\normalsize{$^{2}$ Santa Fe Institute, 1399 Hyde Park Rd, Santa Fe NM 87501 USA} \\
\normalsize{$^{3}$Service Physique Th\'eorique, DSM, C.E. Saclay 91191 Gif/Yvette, France}\\
\\
\normalsize{$^\ast$To whom correspondence should be addressed; E-mail:  asl@lanl.gov}
}

\date{}

\begin{document} 


\baselineskip24pt


\maketitle


\begin{sciabstract}
We present a sequence-based probabilistic formalism that
directly addresses co-operative effects in networks of 
interacting positions in proteins, providing significantly
improved contact prediction, as well as accurate quantitative prediction of
free energy changes due to non-additive effects of multiple mutations.
In addition to these practical considerations, the agreement of our
sequence-based calculations with experimental data for both structure and 
stability demonstrates a strong relation between the statistical 
distribution of protein sequences produced by natural evolutionary processes,
and the thermodynamic stability of the structures to which these sequences fold.
\end{sciabstract}

\vfill
\eject




This manuscript was originally written in 2002 and available from
http://library.lanl.gov/cgi-bin/getfile?01038177.pdf
It's being deposited here for greater ease of access.

Our approach is analogous to solving an inverse problem of statistical mechanics:
determine the physical interaction parameters of a twenty-state spin system 
given a set of sequences drawn from the Boltzmann equilibrium distribution.
The sequences we consider are sets of aligned protein sequences drawn 
from variable sequence families defined in the Pfam database \cite{PFAM}.
We assume that within each family the sequences adopt a common 
(but in principle unknown) fold whose underlying structure is reasonably 
conserved across the family.
Each sequence of length $L$ of a given family can be viewed as a different global 
state of an $L$-site, twenty-state (for twenty amino acids) spin system, with 
spin-spin (i.e. residue-residue) interactions determined by
(1) the (unknown) structure of the associated fold, and (2) the physico-chemical 
characteristics of the residues. Solving the inverse problem to determine 
the underlying physical interactions addresses 
``correlation at a distance'', in which  correlations between locally 
connected sites in an interacting network such as a spin system, or a protein,
can propagate throughout the network, leading 
to observed correlations between sites that have no direct physical 
interaction \cite{Stanley71}. Such propagated correlations 
can be even greater than correlations between any directly connected sites
in the system \cite{Giraud99}.
Previous computational work on abstract models of proteins
\cite{Finkelstein}, as well as a statistical analysis of the frequency 
of ion-pairs in crystal structures of real proteins \cite{Lawrence},
provided early hints that Boltzmann-like statistics are associated 
with aspects of protein architecture. 
In view of complicated evolutionary pressures acting on naturally evolved
protein sequences it is surprising that developing a strictly
thermodynamic approach can, as we demonstrate below, lead to an
accurate predictive methodology for both protein structure and stability.


Other work relating sequence statistics to physical interactions,
but restricted to assuming independent (non-interacting) sites, successfully
characterized protein-DNA binding interactions given 
sequence data 
\cite{VonHippel, Stormo2000, Benos01a}.
``Semi-rational" protein sequence design, see e.g. \cite{Fersht98},
also assumes independent sites, and analyzes 
natural sequence variation to suggest mutations leading to greater 
thermodynamic stability.
However, analysis of mutations in sets of aligned sequences,
first for RNA sequences \cite{Gutell92} and later for protein 
sequences \cite{Korber93}, has shown that mutations in pairs of sequence 
positions are often correlated. 
Such pairwise correlations
have been used in attempts to predict spatially proximate residues (contacts)
in folded proteins
\cite{Shindyalov94, Gobel94, Taylor94,  Neher94, Clarke95, Thomas96, Larson2000}. 
The hypothesis is that pairs of variable residue positions,
possibly distant along the sequence but spatially proximate in 
the folded molecule, will display significant covariation. 
Published approaches analyze correlations between at most two sequence 
positions at a time, hence they inherently assume that each potentially 
interacting pair of positions under consideration is physically isolated from 
all other positions \cite{Mutual}. This assumption is reasonable for RNA molecules, given the 
saturating hydrogen bond interaction between base-pairs, and accuracy of
contact prediction for RNA using pairwise covariation formulae 
is relatively high \cite{Gutell92}. This assumption is not reasonable for
the typically diffuse and networked interactions among amino acids,
and accuracy of contact prediction for proteins using pairwise
covariation formulae is relatively poor.
Pairwise covariation formulae were recently used for a qualitative 
description of stability changes upon mutations in the
SH3 domain, as well as for contact prediction \cite{Larson2000}.
Attempts to chain together separate pairwise analyses 
to approximate interaction networks in proteins \cite{Lockless99} 
can be illuminating, suggesting that a complete formalism to address 
network effects would be fruitful.

The Boltzmann network method presented here does not treat each individual
pair of sites of interest as isolated from other residues. Instead,
we construct a probability distribution describing full length sequences of 
length $L$ for each protein sequence family.
Any given sequence alignment typically contains enough data to estimate
only single and 
pairwise amino acid frequencies with reasonable accuracy. One point of departure from
previous analyses using single and pairwise frequencies is that we adopt an
information theoretic viewpoint, and ask for the least biased
probability distribution, {\it defined over all $L$ sites}, whose first and second order
moments match the single and pairwise amino acid frequencies of the given data.
``Least biased'' is defined to be the maximum entropy distribution \cite{Cover91},
which in our context may be intuitively viewed as the flattest distribution among the many
distributions that have first and second order moments matching the amino acid
frequencies in the given data \cite{InsertBacknote}.

The maximum entropy distribution whose moments match a given set of single and pairwise amino acid 
frequencies may be written in the following form \cite{Lapedes97}, reminiscent of thermal Boltzmann statistics
$$P(X) = \frac {exp\lbrack -E(X) \rbrack} {Z} \, ,  \eqno (1) $$
where $E$ is a sum of single and pairwise interactions among potentially all amino acids
$$ E(X) = \sum_{\alpha \beta i j} \lambda^{\alpha \beta}_{ij} x^{\alpha}_i x^{\beta}_j + \sum_{\alpha i } \lambda^{\alpha}_i x^{\alpha}_i  \, .
\eqno (2)
$$
$x^{\alpha}_i$ denotes the residue present at position $i$ in sequence $X$,
it has the value $1$ if amino acid $\alpha$ is present at sequence position
$i$, and is $0$ otherwise. The $\lambda's$ are adjustable
parameters (to be determined) such that the calculated first and second order moments of this distribution
match the single and pairwise amino acid frequencies in the given sequence alignment.
$i$ and $j$ label sequence positions ($1$ to
$L$), and $\alpha$ and  $\beta$ label the twenty possible amino acids.
$Z$ is a normalization factor.  It can be shown \cite{Backnote4} that matching the 
moments of the maximum entropy distribution to the given sequence data is equivalent
to maximizing the loglikelihood of the given sequence data given the
parametric form, Eqns. (1,2), for the probability distribution.
This formalism is related to 
Boltzmann Machines \cite{BoltzmannMachines} and Graphical 
Models \cite{GM}, used in other contexts.

So far, $E$ is merely
a suggestive symbol appearing in a probability distribution, Eqns. (1,2),
describing
sequence statistics of an alignment. However, it is shown below if the 
$\lambda's$ are adjusted so that the moments of the distribution match the given
amino acid frequencies, then $E$  is highly correlated with a real, physical, thermodynamic free energy of unfolding. Furthermore, we use the 
probability distribution over all $L$ sites, Eqns. (1,2), to resolve issues
of correlation at a distance (network effects) in proteins, resulting in
significantly improved contact prediction from sequence information.

We consider aligned sequences for eleven domains \cite{Backnote1} taken from the
Pfam \cite{PFAM} database, with associated x-ray crystal structures
taken from the Protein Data Bank\cite{PDB}.
These domains were chosen to be diverse in sequence (less than
50\% pairwise sequence identity) and to have more than 200 sequences per 
family.  The distance between a pair of residues was defined to be the distance 
between their carbon $\beta$ atoms,
and pairs of residues with carbon $\beta$ distance of less than 7
{\it A}ngstroms were defined to be in contact (carbon  $\alpha$ coordinates
were used for glycines). Results reported below are robust to 
changes in definition of contact.

Prediction of which residues are directly interacting (i.e. in physical
contact) uses the concept of conditional mutual information \cite{Cover91}
applied to $P(X)$  after the $\lambda's $ have been determined for
each sequence family.
In our context,
conditional mutual information, $CMI$, measures the degree of covariation
between residues at sequence positions $i$ and $j$ 
that is solely due to direct effects of $i$ on $j$ (and {\it vice versa}), factoring out
contributions to the correlation between $i$ and $j$ caused by interaction
of both $i$ and $j$ with the rest of the network of residues. It is a 
discrete (and nonlinear) analogue of linear partial correlation analysis \cite{Parcorr, Afonnikov97}
and is intuitively described by this process:
(a) freeze all residues other than those at $i$ and $j$  to a fixed state,
thus preventing information propagation through the rest of the network,
(b) calculate the mutual information between $i$ and $j$, using $P(X)$, above,
with the rest of the network frozen, and (c) average this result over all
possible frozen states of the rest of the network \cite{Backnote5}.
Pairs of sites with high $CMI$ (over a user defined threshold)
are predicted to be in contact.

Two quantities, specificity and sensitivity, are typically used to
characterize predictive ability.
Specificity is defined as the fraction of predicted contacts that are 
actual contacts (as defined by carbon $\beta$ distances) i.e. 
the overall probability that a predicted contact is correct.
Sensitivity is defined as the fraction of actual contacts that 
are correctly predicted. High specificity is more desirable than
high sensitivity, because in our context predicting even a small number 
of contacts with high accuracy provides extremely valuable constraints on
{\it ab initio} protein structure calculations \cite{Saitoh93, Skolnick98a}. Hereafter we refer to specificity as ``accuracy''.
To survey accuracy as a function of $CMI$ threshold
we successively lowered the $CMI$ threshold, in effect walking down
a list of predicted contact pairs ordered by $CMI$ value, for each domain.
This process yields accuracy of prediction as a function of the
number of pairs predicted to be in contact \cite{Backnote6}.

To compare our method  to others we also analyzed contact prediction
accuracy using 
(a) a pairwise covariation measure \cite{Larson2000}
(denoted as $\Phi AM$ for 
$\Phi$ Association Method \cite{Backnote7}, which we believe to be the most accurate of published methods),
(b) conventional pairwise mutual information \cite{Mutual} (denoted as $MI$) 
and (c) a baseline reference resulting from random selection of 
position pairs (denoted as $Random$).
The measure
used in (a) above also incorporates some correction for phylogenetic artifacts.
Fig. (1) shows overlaid curves for accuracy of contact prediction 
via the different methods, versus number of predicted contacts,
for the SH3 domain. The most accurate method is the Boltzmann network
method, which uses conditional mutual information to predict contact pairs.
\begin{figure}[htbp]
\begin{center}
\leavevmode
\rotate[r]{\epsfxsize=4.0in \epsffile{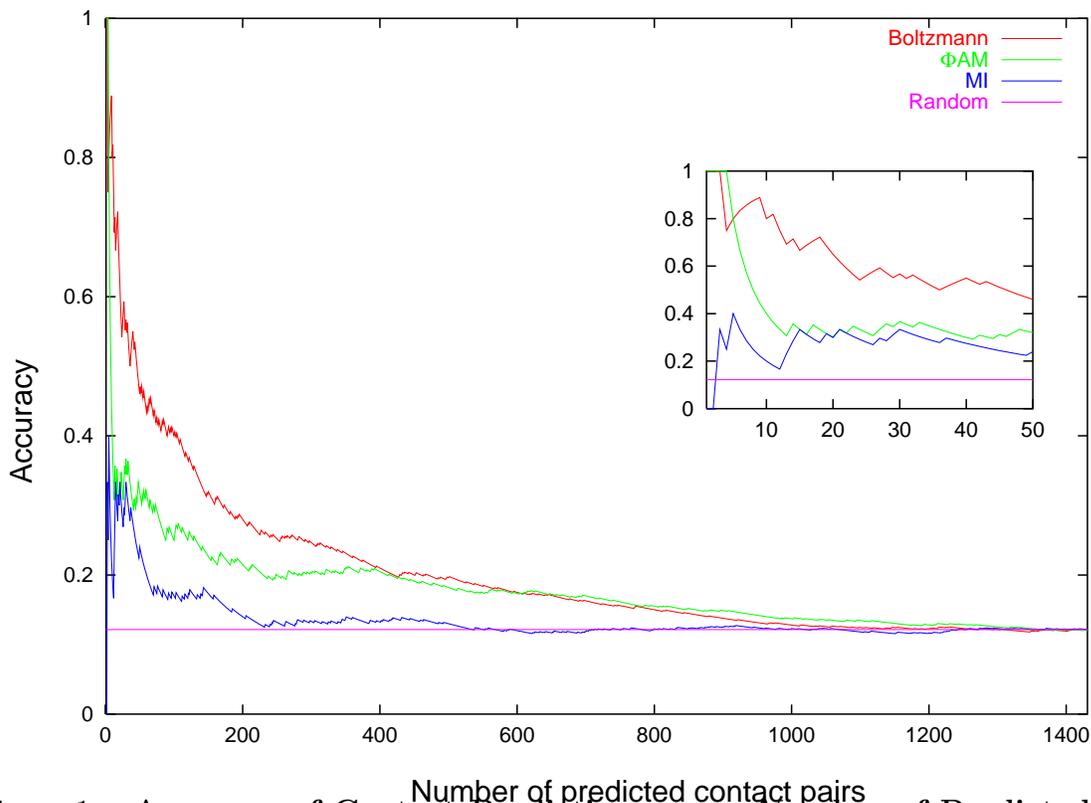}}
\caption{ {\bf Accuracy of Contact Prediction versus Number of Predicted Contacts:} 
Accuracy of prediction (y-axis) vs. number of predicted
contact pairs (x-axis) for the SH3 domain is shown.
$Boltzmann$, the top curve,  is the
result of the Boltzmann network method.
$\Phi AM$ is the result of what we believe to be 
the most accurate published pairwise covariation method \cite{Larson2000} 
(does not address network interactions, does address
phylogenetic artifacts), $MI$ is the result of pairwise mutual  information 
(does not address network interactions),
and $Random$ is the average result of 
picking at random a specified number of contacts.
The inset blows up the region from $1$ to $50$ predicted contacts.
The accuracy of contact prediction using the Boltzmann network method, which
incorporates co-operative effects among residues, significantly exceeds that 
of other methods.
}
\end{center}
\end{figure}
%
Accuracy varies somewhat from family to family, therefore we show
the averaged accuracy over eleven domains in Fig. (2) using the same
four predictive methods. The Boltzmann network method
has on average
consistently higher accuracy for a greater number of predicted contacts.
Predicted contacts for the eleven domains using the Boltzmann network method
are available in the supplemental material \cite{SupplementalMaterial}.


\begin{figure}[htbp]
\begin{center}
\leavevmode
\rotate[r]{\epsfxsize=4.0in \epsffile{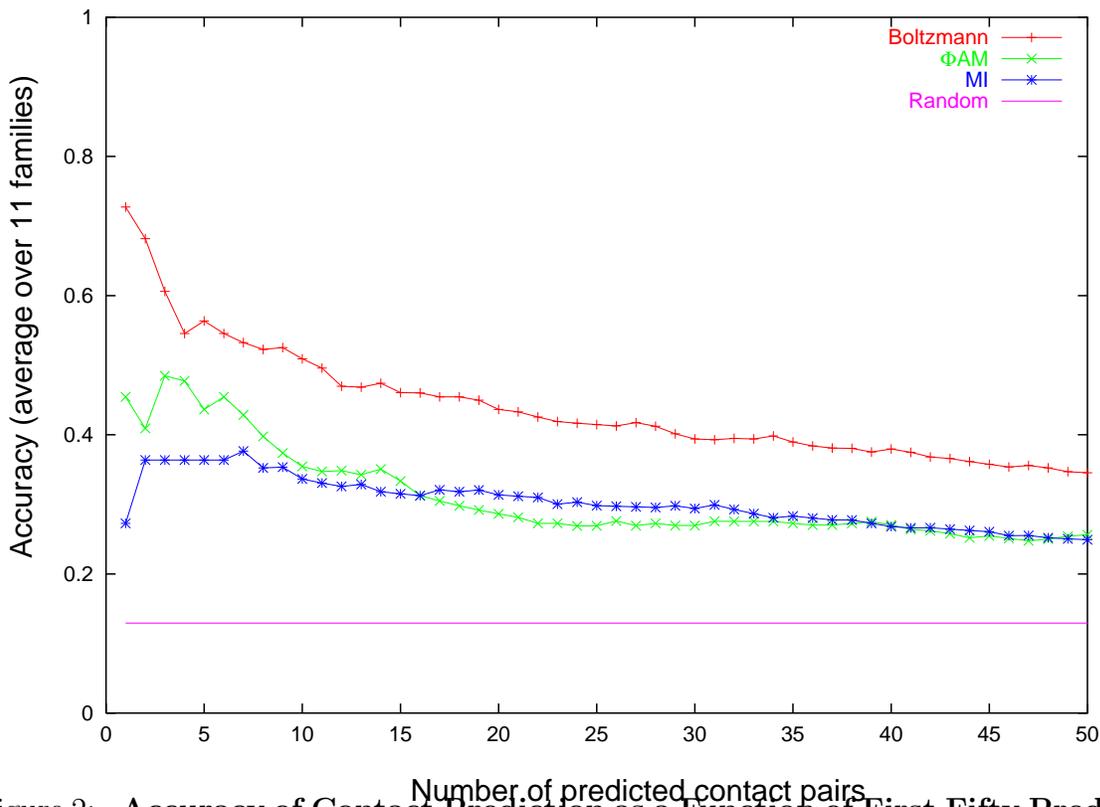}}

\caption{ {\bf Accuracy of Contact Prediction as a Function of First Fifty Predicted Contacts:}
An Average over eleven domain families, for the same predictive methods of Fig. (1).
$Boltzmann$, the top curve, is the result of the  Boltzmann network method
presented here and has significantly higher average accuracy, demonstrating
the importance of addressing co-operative effects within proteins.
}
\end{center}
\end{figure}

The maximum entropy probability  distribution, Eqn. (1), has a thermal, Boltzmann
form with exponent $E(X)$. After the $\lambda's$ have been determined for a
given sequence alignment, $E(X)$ assigns an ``energy'' value to any 
sequence $X$. Interpreting $E(X)$ as an effective free 
energy relative to the unfolded state allows a free energy of unfolding \cite{Fersht90},
$\Delta G= -E(X)$, to be predicted using our formalism.  Changes in sequence, $X$,
will change $E(X)$ and hence the $\Delta G$ 
of sequence mutants can be calculated and compared to experiment. Experimentally determined
melting temperatures (assumed proportional to the free energy of unfolding)
for wildtype Fyn SH3 sequence, and  for
a set of single, double and triple mutants of the wildtype
were reported in \cite{Larson2000}.
To assess how well co-operative non-additive effects are captured by our formalism,
we calculated $\Delta G$ values after the $\lambda's$ had been determined in two different ways: 
(1) the interaction parameters, $\lambda^{\alpha \beta}_{ij}$,
were allowed to adjust during the determination of the probability distribution
$P(X)$, (2) the 
interaction parameters, $\lambda^{\alpha \beta}_{ij}$, 
were held fixed to zero, allowing only additive
effects to be captured by the remaining adjustable single site $\lambda^{\alpha}_{i}$ 
parameters \cite{Backnote8}. As will be seen below, the correct
prediction of the effects of even single site mutants requires consideration
of the other sites with which it interacts. The eleven residues identified 
by a structural analysis \cite{Larson2000} to be in the hydrophobic core 
of the SH3 domain were selected for use in assessing $\Delta G$ prediction,
i.e. the $\lambda$ parameters used for computing $\Delta G$
allowed potential  interaction among all eleven sites of the hydrophobic core.
Significant sequence variation is 
necessary input information for our method, and so within this set of eleven 
core positions we report $\Delta G$  values for mutations involving the 
three positions (26, 39 and 50 in the numbering scheme of \cite{Larson2000})
that displayed the highest mutual information.

Experimentally determined melting temperatures were reported \cite{Larson2000}
for four single, four double, and three triple mutants, in addition to 
the wildtype for these three positions.
In Fig. (3) the  difference of the mutant and wildtype 
$\Delta G$'s as computed by our method for these mutant domains
is shown to be highly correlated (absolute value of correlation $0.91$) with the experimentally measured 
melting temperatures. If non-additive and co-operative effects are
disallowed by holding the interaction terms to zero  then the correlation is
poor (absolute value of correlation $0.02$) and the signs of the predicted 
$\Delta G$ are incorrect, Fig. (4). 

\begin{figure}[htbp]
\begin{center}
\leavevmode
\rotate[r]{\epsfxsize=4in \epsffile{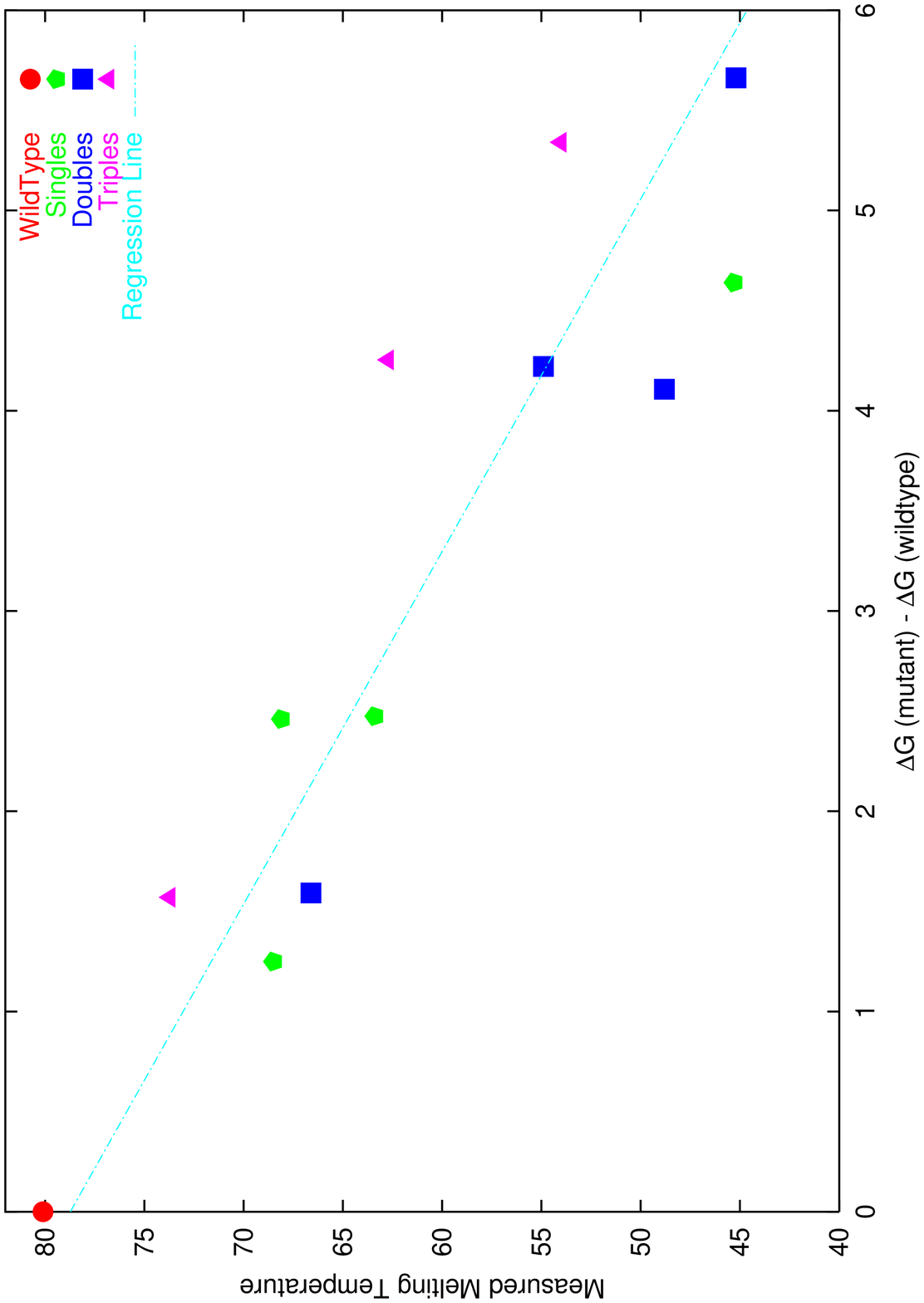}}

\caption{{\bf Measured Melting Temperature versus Predicted Free Energy of Unfolding: Network
Interactions Allowed} The free energy of unfolding, $\Delta G$,
computed using the Boltzmann network method
versus experimentally measured melting temperature for eleven
mutants of the SH3 domain (four singles, four doubles, three triples) as 
well as the the wildtype. Co-operative and non-additive effects were
allowed, resulting in a good correlation of computation with experiment
(absolute value of the correlation is $0.91$).
The single site mutant, $I50F$, as discussed by Larson,
involves mutating to a residue more frequent in the alignment and yet
is measured to be quite destabilizing with a measured melting temperature of
$45.3$. Only if network interactions are allowed is this
single site mutant correctly predicted as quite destabilizing.
The triple mutant $F26I/A39G/I50F$, with a measured melting temperature
of $73.7$, involves $I50F$ with additional compensatory second site mutations. 
It is correctly predicted as just mildly destabilizing compared to wildtype only if
network interactions are allowed. Comparison of this figure (network
interactions included) to Fig. (4) (network interactions excluded),
shows in general the importance of network interactions.
}
\end{center}
\end{figure}

\begin{figure}[htbp]
\begin{center}
\leavevmode
\rotate[r]{\epsfxsize=4in \epsffile{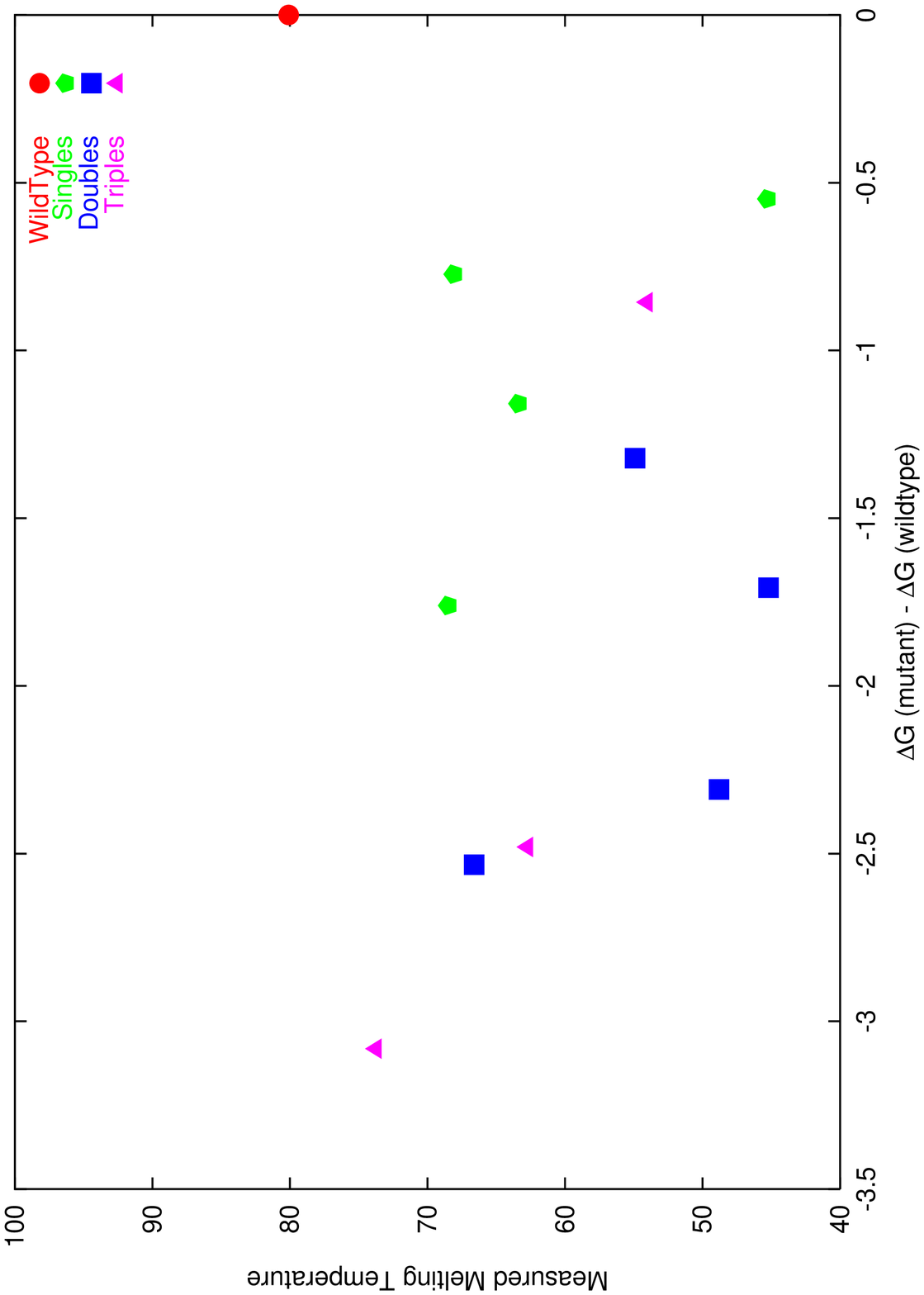}}
\caption{ {\bf Measured Melting Temperature versus Predicted Free Energy of Unfolding: Network Interactions Not Allowed}
The free energy of unfolding, $\Delta G$,
versus experimentally measured melting temperature for the same
eleven mutants of the SH3 domain and for wildtype, as Fig. (3), but
when co-operative and non-additive effects are disallowed by holding the interaction
parameters, $\lambda_{ij}^{\alpha \beta}$, to zero.
There is a dramatic decrease in correlation of computation with experiment
(absolute value of the correlation is now $0.01$), and even the signs of the 
stability changes are incorrect when network interactions are disallowed.
}
\end{center}
\end{figure}

The high correlation of predicted and measured $\Delta G$ shown in
Fig. (3) suggests performing a computational survey of all 
possible ${(20^3)}$ mutations to search for other interesting single,
double and triple mutants.
We found no statistically significant prediction of a mutant considerably
more stable than wildtype, although the single site mutant F26L has
a predicted melting temperature similar to wildtype with a value of $84.9$.
Predictions of melting temperature for other interesting three site
mutants are available in the supplemental material 
\cite{SupplementalMaterial}. Results 
of an extensive computational search among all $(20^{11})$ possible sequences
defining a total redesign of the eleven site
hydrophobic core of the SH3 domain are also presented 
\cite{SupplementalMaterial}.

%


The success of the Boltzmann network formalism in predicting
free energy changes upon mutation clearly demonstrates a deep 
relationship between the statistics of sequences selected by
natural evolutionary processes and the thermal
stability of the structures to which these sequences fold. 
However, such a strong relationship would not necessarily be expected 
given that protein sequences produced by evolution are
strongly affected by functional constraints in addition to stability constraints 
\cite{Fersht99}. A possible explanation of the statistics-stability
relation is that functional properties are
typically confined to localized regions of a protein,
e.g. binding sites, and that optimization of small local regions for
functional fitness occurs after global selection for sequences that stably fold.
An independent, computational investigation of the extent 
to which sequences are shaped by natural selection for stability
was published recently \cite{Baker2000} 
although contact prediction and prediction of free energy changes was not
explicitly addressed. In contrast to our sequence based approach, 
this work used structural information,
combined with an all-atom free energy function incorporating a variety of physical effects
to computationally design sequences for a variety of domains. 
The native, naturally occurring sequence for each structure 
considered was found to be close to optimal for each structure, and for SH3,
the pairwise correlations between sites in
a set of computationally designed sequences recapitulated the correlations 
observed in a set of native SH3 core sequences. 
The extent to which the Boltzmann network's energy function, $E(X)$, involving
empirical parameters determined solely from sequence information for
each domain family, can be identified with the physical/structural
effects defining the energy function of this structure-based
complementary study remains an interesting issue.

Limiting factors in application of the Boltzmann network algorithm include
(1) the amount of naturally evolved sequence data currently available 
per family (size of the sequence alignment), and (2) the phylogenetic
relatedness (and associated selection artifacts) of these sequences.
Modifications to the algorithm presented here, e.g. (1) consideration 
of statistical significance of the fitted $\lambda$ parameters,
and (2) addressing phylogenetic relationships of sequences in
an alignment, have the potential to further increase accuracy using 
naturally evolved sequence sets.

However, the ability to create in the laboratory
totally novel sequences for protein domains via artificial evolution
techniques such as phage display \cite{Baker97} \cite{Szostak2001}, 
promises new, rich, and diverse sequence sets with well characterized 
{\it in vitro} selection pressures.  Such sequence data, when available 
in greater quantity, and analyzed with the methods herein, offer
a new paradigm for sequence based structure prediction, and for the
computational design of sequences with preferred properties.

\clearpage


\vfill \eject

{\bf Appendix: Supporting Online Material}

\section{Determining the $\lambda$ Parameters by Maximum Likelihood Analysis}

The maximum entropy probability distribution subject to
constraints on the first and second order moments, has an exponential form
$$P(X) = \frac {exp-\lbrack E(X)\rbrack}{Z} \, ,
$$
where $E$ is a sum of single and pairwise interactions,
$$ E(X) = \sum_{\alpha \beta i j} \lambda^{\alpha \beta}_{ij} X^{\alpha}_i X^{\beta}_j + \sum_{\alpha i} \lambda^{\alpha}_i X^{\alpha}_i
$$
and
$$Z = \sum_{X} exp-\lbrack E(X) \rbrack$$ is a sum over all possible $(20^L)$ sequences of
length $L$ which normalizes the distribution [1]. The $\lambda's$ are Lagrange
multipliers implementing the constraints that the first and second order
moments of the distribution match the single and pairwise amino
acid frequencies in a given sequence alignment. Each sequence
$X$ of the alignment may therefore be assigned a probability, $P(X)$,
which is a function of the $\lambda's$.

For each sequence alignment considered one may write
the joint probability of all $S$ sequences of the alignment as
a function of the $\lambda 's $ (assuming that the sequences
are independent)  as
$$ P(Sequences) = \prod_{s=1}^{s=S} \frac {exp-\lbrack E(X(s))\rbrack} {Z} $$
where $s$ references each sequence of the alignment.
Although naturally evolved sequences that
are related by a phylogenetic tree are not independent, making the
assumption of independence, for simplicity, still yields results 
of high accuracy (this assumption of sequence independence is of course
unrelated to issues of site independence within a sequence). 
Properly addressing the phylogenetic relatedness of sequences
is complicated, but has the potential to increase accuracy still further.

Taking $logs$ of both sides yields
$$ log \lbrack P(Sequences) \rbrack  = {-\lbrack \sum_{\alpha \beta i j s} \lambda^{\alpha \beta}_{ij} X{^{\alpha}_i}(s) X{^{\beta}_j}(s) + \sum_{\alpha i s} \lambda^{\alpha}_i X{^{\alpha}_i}(s) + {S*log(Z)} \rbrack}
$$
$$  = -S* \lbrack \sum_{\alpha \beta i j}  \lambda^{\alpha \beta}_{ij} \overline{X^{\alpha}_i X^{\beta}_{j}}  + \sum_{\alpha i } \lambda^{\alpha}_i * \overline{ X^{\alpha}_i} + log(Z) \rbrack
$$
Here, $\overline{ X^{\alpha}_i}$ and $\overline{X^{\alpha}_i X^{\beta}_{j}}$
represent, respectively, the single and pairwise amino acid frequencies
obtained by simple counting in the given aligned sequence data set.

A steepest ascent step, maximizing $log \lbrack P(Sequences) \rbrack$, changes the $\lambda$
parameters by an amount proportional to the gradient of $log \lbrack P(Sequences) \rbrack$
with respect to the $\lambda 's$
$$ \Delta \lambda^{\alpha \beta}_{ij} \sim \frac {\partial log \lbrack P(Sequences) \rbrack}{\partial \lambda^{\alpha \beta}_{ij}} \sim  (\overline{X^{\alpha}_i X^{\beta}_{j}}- <X^{\alpha}_i X^{\beta}_{j}>)$$
$$ \Delta \lambda^{\alpha }_{i} \sim \frac {\partial log \lbrack P(Sequences) \rbrack}{\partial \lambda^{\alpha }_{i}} \sim  (\overline{X^{\alpha}_i }- <X^{\alpha}_i >)$$
where
$$ <X^{\alpha}_i X^{\beta}_{j}>= -\frac {\partial log(Z)}{\partial \lambda^{\alpha \beta}_{ij}} $$
$$ <X^{\alpha}_i >= -\frac {\partial log(Z)}{\partial \lambda^{\alpha }_{i}} $$
represent the second and first order moments of the distribution, respectively.
In principle,
evaluating these moments involves $({20}^L)$ summations, however since they
are simple averages 
they may be efficiently estimated in practice via Monte Carlo [2]. Once the
moments have been estimated at a current setting of the $\lambda's$, the $\lambda's$ are changed by an amount
proportional to $\Delta \lambda$ and the process is iterated to
convergence. This procedure is
essentially the ``training'' algorithm for a Boltzmann machine [3]
when there are no hidden units. Note that at the maximum 
of $log \lbrack P(Sequences) \rbrack$, i.e., when $\Delta \lambda^{\alpha \beta}_{ij}=0$, 
$\Delta \lambda^{\alpha }_{i}=0$,
the single and pairwise moments of the distribution match
the single and pairwise amino acid frequencies of the given 
sequence alignment. 
Furthermore, it is possible to show by consideration of mixed
second derivatives, such as
$\frac { {\partial}^2 log \lbrack P(Sequence) \rbrack}{\partial \lambda^{\alpha \beta}_{ij} \partial \lambda^{\gamma \delta}_{kl} }$, 
$\frac { {\partial}^2 log \lbrack P(Sequence) \rbrack}{\partial \lambda^{\alpha \beta}_{ij} \partial \lambda^{\gamma}_{k}}$  that $log {\lbrack P(Sequence) \rbrack}$ is a convex function
and that there are no local maxima.

For the results reported in the main text,
the $\lambda's$ were initialized with the $\lambda^{\alpha \beta}_{ij}$
interaction terms set to zero, and the $\lambda^{\alpha}_{i} $ terms
chosen to match the single site amino acid frequencies of each given
sequence alignment. 
To evaluate moments of the distribution given some current values of 
the $\lambda's$,
$400,000$ sequence configurations
were obtained by generating a Monte Carlo chain of
$4,000,000$ steps, and keeping every tenth configuration
of this chain  when estimating the $<>$ moments.
Change in the $\lambda's$,
$\Delta \lambda$, are zero, and the iterative process
converges, when the moments exactly match the amino acid frequencies.
This occurs when the likelihood is a maximum and the gradient is zero.
Effective convergence was reached in (very roughly)  on the order of
$10,000-15,000$ changes of $\Delta \lambda's$ or $40-60$
hours of computer time (depending on domain size), on a dedicated
single processor 1 $ghz$ cpu with $500$ megbytes of memory. No significant
effort was made to optimize code beyond addressing the most
obvious inefficiencies.

\section {Predicted Contacts for Eleven Families}

The top $50$ predicted contact pairs, using the Boltzmann network
method (see main text), for each of the $11$ Pfam families follows.
Each column, representing one protein family, is ordered by descending value
of conditional mutual information. The numbering scheme for specifying 
position pairs of each predicted contact uses the residue number appearing
on the ``ATOM'' lines in the PDB files listed at the top of each column.

$$
\begin{array}{ccccccccccc}
1ubi&1fjl&1shg&3icb&6pti&1awc&1mdy&1sha&5fd1&1a17&1be9 \\
\hline
19,24&6,13&44,53&56,58&20,44&72,101&139,141&6,33&24,25&34,50&367,375 \\
19,57&24,29&31,44&50,66&24,31&79,82&125,135&18,31&3,4&43,46&329,33 \\
24,57&3,7&25,31&49,50&13,36&94,95&129,155&56,80&4,7&35,51&336,37 \\
20,22&17,52&31,53&49,66&21,48&91,97&136,148&33,86&12,17&31,50&353,39 \\
28,57&6,7&53,58&45,70&21,31&96,98&120,125&65,86&17,18&53,56&366,36 \\
16,22&28,30&17,24&58,60&7,42&80,101&119,149&79,82&7,10&34,38&346,35 \\
20,25&29,43&21,22&50,53&27,52&77,81&141,161&56,84&6,7&32,36&319,32 \\
23,26&2,7&25,53&63,66&14,38&72,80&125,151&11,45&5,22&46,49&316,35 \\
19,25&27,29&35,43&50,52&9,16&82,83&123,129&10,14&1,7&42,44&350,35 \\
20,26&24,45&15,55&46,58&48,52&79,80&125,148&80,84&1,10&29,33&325,37 \\
23,31&29,36&42,55&58,67&23,54&77,80&135,155&6,28&22,23&38,46&366,36 \\
16,24&31,55&15,21&48,70&21,52&80,83&151,153&31,84&13,18&33,36&323,36 \\
22,24&8,43&44,58&62,63&27,54&81,83&124,136&7,28&7,18&58,61&338,35 \\
19,20&27,30&14,26&50,70&44,52&97,98&116,131&59,60&1,4&36,51&323,37 \\
23,25&10,13&15,42&69,70&15,21&80,81&129,145&14,61&3,10&39,51&322,38 \\
24,28&28,34&9,33&45,73&8,25&79,87&125,147&42,46&5,9&58,60&335,33 \\
20,24&21,24&13,57&60,62&27,48&88,92&123,125&49,86&1,26&59,61&341,35 \\
19,28&23,27&24,25&58,64&22,31&83,86&147,161&84,85&1,22&38,43&369,37 \\
16,18&28,33&13,56&66,67&46,48&99,101&139,142&79,86&5,7&41,43&338,34 \\
16,56&8,58&33,39&53,69&36,39&81,87&152,154&6,10&10,13&37,40&358,39 \\
16,20&34,39&25,52&49,53&16,34&89,93&110,114&31,33&9,13&33,51&330,371\\
26,28&29,45&10,29&63,67&23,24&83,87&116,124&28,80&2,22&37,41&345,350\\
25,54&33,34&17,33&53,66&22,53&72,86&120,136&6,55&5,26&35,50&316,325\\
22,56&30,43&17,34&53,63&21,24&77,101&122,125&14,43&1,3&32,33&367,370\\
3,18&15,45&12,13&46,47&7,11&79,97&145,157&10,36&1,2&39,44&323,346\\
54,64&36,39&34,45&46,67&17,36&72,77&136,140&36,79&2,4&58,59&314,325\\
25,26&4,7&35,37&62,69&16,36&77,83&113,116&28,46&10,22&48,52&314,338\\
20,54&21,34&7,36&62,67&9,34&72,82&131,141&25,28&2,18&29,30&361,367\\
18,22&28,54&39,52&58,72&6,23&93,101&113,136&10,78&5,10&54,57&322,327\\
12,25&29,39&10,59&57,70&10,13&72,83&153,157&10,28&10,26&30,36&362,376\\
23,28&7,10&38,52&56,57&8,26&85,88&116,138&7,10&17,25&28,61&329,371\\
3,16&15,17&22,52&46,72&8,46&97,101&144,152&9,33&19,22&28,32&379,380\\
19,54&29,34&10,52&50,63&27,31&97,100&124,145&9,11&3,5&38,50&330,372\\
12,20&23,34&33,37&45,48&23,26&86,87&147,152&78,82&4,5&32,51&367,378\\
16,57&7,41&31,52&58,62&29,49&77,97&114,122&14,74&7,13&28,54&337,382\\
19,22&7,13&12,39&56,61&44,46&77,82&134,138&61,74&1,23&33,37&372,376\\
60,64&34,45&22,53&56,62&28,29&70,80&128,133&41,82&1,13&33,61&325,337\\
23,39&41,43&9,60&49,73&8,16&93,100&148,151&47,61&1,5&32,61&362,386\\
22,25&8,14&33,34&46,53&6,28&79,81&136,159&11,46&2,13&48,57&322,345\\
54,60&26,44&10,30&49,52&15,36&70,72&125,154&36,41&5,13&50,53&341,348\\
20,23&32,36&9,18&66,73&16,28&87,89&110,149&11,59&4,6&31,53&354,388\\
23,32&26,46&9,17&67,71&10,42&80,97&136,151&31,40&6,10&32,55&314,350\\
25,28&2,33&11,25&53,70&21,39&70,85&113,120&10,60&24,26&30,33&361,375\\
18,26&24,39&24,57&49,61&31,48&74,80&116,127&37,41&25,26&49,53&345,346\\
16,26&4,6&9,42&52,70&21,22&82,84&148,154&73,75&13,19&28,30&341,359\\
24,56&19,30&27,38&52,62&7,50&72,74&125,142&25,84&4,18&30,40&314,353\\
23,53&18,19&17,21&52,53&52,54&98,99&135,139&24,28&19,25&40,45&323,326\\
26,31&22,23&33,55&51,62&8,10&96,100&142,145&10,45&18,19&30,37&343,387\\
19,64&4,10&42,59&47,66&10,48&82,102&134,142&45,60&5,17&39,50&361,389\\
12,19&22,28&9,43&66,70&21,27&70,82&120,155&31,46&2,5&44,48&341,354
\end{array} 
$$


\section {A Computational Survey of ${(20^3)}$ Mutants of the Hydrophobic Core of Fyn SH3}
The high correlation reported in the main text for calculated $\Delta G$
values with measured melting temperatures, for mutations in positions
26, 39 and 50 (numbering scheme of reference [4]),
suggests performing a computational survey of 
all $(20^3)$ possible mutants in these positions.
Using the regression line of Fig. (3) of the main text enables conversion of
any calculated $\Delta G$ to a predicted melting temperature.
We surveyed all ${(20^3)}$ possible mutants in these three positions and
selected those mutant sequences with predicted melting temperatures within 
the range of measured melting temperatures of Fig.(3), in effect
interpolating  new sequences between existing sequences with
measured melting temperatures. Regarding sequences outside of this range:
sequences with significantly higher melting temperatures were not found;
on the other end of the temperature range, sequences utilizing amino 
acids that were rare in the initial sequence alignment depend on $\lambda$ parameters that are poorly determined, and were eliminated from the set of 
significant predictions. $50$ such triple mutants, ordered by predicted melting temperature,
are listed below. The numbering scheme is that of reference [4]: residues listed
correspond to positions 4,6,10,18,20,26,28,37,39,50,55.


\small
$$
\begin{array}{ll}

Sequence&Temperature\\
\hline

FAYLFLIWAIV&84.9\\

FAYLFFIWAIV&78.7\\

FAYLFIIWAIV&71.6\\

FAYLFIIWGFV&69.8\\

FAYLFIIWGIV&69.6\\

FAYLFLIWAVV&69.3\\

FAYLFMIWAIV&69.2\\

FAYLFLIWGIV&67.7\\

FAYLFFIWGIV&64.7\\

FAYLFLIWVIV&63.5\\

FAYLFFIWAVV&60.8\\

FAYLFLIWCIV&60.7\\

FAYLFFIWVIV&60.1\\

FAYLFLIWALV&59.6\\

FAYLFIIWGVV&58.0\\

FAYLFIIWGLV&57.8\\

FAYLFCIWAIV&57.6\\

FAYLFIIWAVV&57.3\\

FAYLFLIWAFV&56.6\\

FAYLFMIWGIV&56.3\\

FAYLFFIWCIV&56.3\\

FAYLFLIWIIV&56.0\\

FAYLFLIWGFV&55.7\\

FAYLFIIWAFV&55.3\\

FAYLFIIWVIV&55.0\\

FAYLFLIWGVV&54.8\\

FAYLFLIWVVV&54.7\\

FAYLFFIWGFV&54.7\\

FAYLFLIWAYV&54.7\\

FAYLFMIWAVV&54.5\\

FAYLFLIWAAV&54.1\\

FAYLFFIWALV&54.1\\

FAYLFIIWCIV&53.2\\

FAYLFIIWALV&52.8\\

FAYLFIIWGYV&52.8\\

FAYLFAIWAIV&52.7\\

FAYLFFIWIIV&52.4\\

FAYLFFIWAFV&52.3\\

FAYLFLIWCVV&52.0\\

FAYLFMIWVIV&50.7\\

FAYLFIIWGAV&50.7\\

FAYLFIIWIIV&49.6\\

FAYLFSIWAIV&49.6\\

FAYLFFIWGVV&49.5\\

FAYLFMIWCIV&49.5\\

FAYLFLIWGLV&49.4\\

FAYLFFIWVVV&49.0\\

FAYLFFIWAYV&48.8\\

FAYLFMIWGFV&48.3\\

FAYLFFIWAAV&48.2\\
\end{array}
$$

\normalsize
\section {A Computational Survey of ${({20}^{11})}$ Hydrophobic Core Sequences of Fyn SH3}
A computational survey of sequence space can also be performed using
the Boltzmann network formalism, even when the number
of potential sequences in the survey precludes exhaustive enumeration.
We illustrate this by suggesting complete redesigns for 
the eleven residue hydrophobic core sequence of SH3, for which an exhaustive
survey of ${({20}^{11})}$ possible core sequences is infeasible.
A stochastic search via simulated annealing, using the modified Lam
schedule for temperature changes [5,6], was used to compile a list
of the $50$ most stable sequences identified during the annealing
process. Of these predicted core sequences, $26$ occur in the initial
sequence alignment, i.e. occur in naturally evolved proteins, and constitute
predictions of the melting temperatures of these natural sequences.
The remaining $24$ sequences constitute predictions of new stable core sequences.
Residues listed below correspond to positions 4,6,10,18,20,26,28,37,39,50,55
in the numbering scheme of reference [4].


\bigskip

\small
$$
\begin{array}{ll}

Sequence&Temperature\\
\hline

FAYLFLIWAIV&84.9\\

VAYLFIVWGFV&84.6\\

VAYLFLIWAIV&79.3\\

VAYLFLVWAIV&79.0\\

FAYLFFIWAIV&78.7\\

AAYLFIVWGFV&77.6\\

VAFLFIVWGFV&77.3\\

FAYLFLVWAIV&76.8\\

VAYLFIIWGFV&76.7\\

VAYLFILWGFV&76.3\\

FAFLFLIWAIV&76.3\\

AAFLFIIWGFV&76.0\\

AAYLFIIWGFV&75.9\\

VAYLLIVWGFV&75.7\\

YAYLFIVWGFV&75.4\\

VAYLFINWGFV&75.4\\

AAFLFIVWGFV&75.2\\

AAYLFLIWAIV&74.0\\

AAFLFIVYGFV&73.6\\

VAYLFIVLGFV&73.3\\

VAYIFIVWGFV&73.1\\

VAYLFIVWGIV&72.8\\

AAFLFIIYGFV&72.7\\

VAYLFIVYGFV&72.6\\

VAYLLVVWGFV&72.3\\

VAYLFIIWGIV&72.0\\

VAFLFLIWAIV&71.9\\

VAFLFIIWGFV&71.9\\

VAFLFIVYGFV&71.8\\

FAYLFIIWAIV&71.6\\

AAYLFINWGFV&71.5\\

AAFLFLIWAIV&71.5\\

VAFLFILWGFV&71.5\\

AAFLFILWGFV&71.3\\

AAYLFILWGFV&71.3\\

FAYLFLIWAII&71.0\\

AAFLFILYGFV&71.0\\

VAYLFFIWAIV&71.0\\

VAYLFLVWGIV&70.8\\

VAYLFLVWGFV&70.4\\

CAYLFIVWGFV&70.2\\

VAYLFVVWGFV&70.2\\

AAYLLIVWGFV&70.2\\

FAYLFLIWAIL&70.1\\

AAFLLIIWGFV&70.0\\

CAYLFLIWAIV&69.9\\

FAYLFIVWGFV&69.9\\

FAYLFIIWGFV&69.8\\

VAYLLVVLGFV&69.8\\

VAYLFLVWAVV&69.7\\
\end{array}
$$

\normalsize

{\bf Supplemental Material References}

[1] A. Lapedes, B. Giraud, L. Liu, G. Stormo,
``Correlated Mutations In Models of Protein Sequences: Phylogenetic and Structural Effects'',
{\it Proceedings of the IMS/AMS 1997 International Conference on Statistics
in Computational Molecular Biology (Seattle 1997)}, published as
Vol 33, Monograph Series of the Inst. for Mathematical Statistics (Hayward Press
)

[2] N. Metropolis, A. Rosenbluth, M. Rosenbluth, A. Teller, E. Teller,
{\it J. Chem Phys.} {\bf 21} p. 1087 (1953)

[3] G. Hinton, T. Sejnowski,
``Learning and Relearning in Boltzmann Machines''
in {\it Parallel Distributed Processing}, eds. D. Rumelhart, J. McClelland 
MIT Press, Cambridge MA (1986)

[4] S. Larson, A. Di Nardo, A. Davidson,
{\it J. Mol. Biol.} {\bf 303} pp. 433-446 (2000)

[5] J.Lam, J. Delosme ``Performance of a New Annealing Schedule'',
in {\it Proc. of the ${25}^{th}$ ACM/IEEE Design Automation Conference},
pp. 306-311, IEEE Computer Society Press (1988)

[6] W. Swartz, C. Sechen `'New Algorithms for the Placement and Routing
of Macro Cells'', in {\it Proc. of the IEEE Intl. Conference on Computer
Aided Design}, pp. 336-339, IEEE Computer Society Press (1990)


\begin{thebibliography}{99}

\bibitem[1]{PFAM}
E. Sonnhammer, S. Eddy , R. Durbin,
{\it Proteins} {\bf 28} pp. 405-420 (1997)

\bibitem[2]{Stanley71} H. Stanley,
{\it Introduction to Phase Transitions and Critical Phenomena}
The International Series of Monographs on Physics
Oxford University Press Inc.  Oxford and New York (1971)

\bibitem[3] {Giraud99} B. Giraud, J. Heumann, A. Lapedes,
{\it Phys. Rev. E.} {\bf 59}, p. 4983 (1999)

\bibitem[4] {Finkelstein} A. Finkelstein, A. Badretdinov, A. Gutin,
{\it Proteins} {\bf 23} pp. 142-150 (1995)

\bibitem[5] {Lawrence} S. Bryant, C. Lawrence,
{\it Proteins} {\bf 9} pp. 108-119 (1991)

\bibitem[6]{VonHippel} O. Berg, P. von Hippel,
{\it J. Mol. Biol.} {\bf 193} pp. 723-750 (1987)

\bibitem[7]{Stormo2000} G. Stormo,
{\it Bioinformatics} {\bf 16} pp. 16-23 (2000)

\bibitem[8]{Benos01a} P. Benos, A. Lapedes, G. Stormo,
{\it Bioessays} {\bf 24} pp. 466-475 (2002)

\bibitem[9]{Fersht98} P. Nikolova, J. Henckel, D. Lane, A. Fersht,
 {\it Proc. Natl. Acad. Sci. USA} {\bf 95} pp. 14675-14680 (1998)

 \bibitem[10]{Gutell92} R. Gutell, A. Power, G. Hertz, E. Putz and G. Stormo,
  {\it Nucl. Acids Res.} {\bf 20} pp. 5785-5795 (1992)

\bibitem[11] {Korber93} B. Korber, R. Farber, D. Wolpert, A. Lapedes,
{\it Proc. Natl. Acad. Sci. USA} {\bf 90} pp. 7176-7180 (1993)


\bibitem[12]{Shindyalov94} I. Shindyalov, N. Kolchanov, C. Sander,
{\it Prot. Eng.} {\bf 7} pp. 349-358 (1994)

\bibitem[13]{Gobel94} U. Gobel, C. Sander, R. Schneider, A. Valencia,
{\it Proteins} {\bf 18} pp. 309-317 (1994)

\bibitem[14]{Taylor94} W. Taylor, K. Hatrick,
{\it Prot. Eng.} {\bf 7} pp. 341-348 (1994)

\bibitem[15]{Neher94} E. Nehe,
{\it Proc. Natl. Acad. Sci. USA} {\bf 91} pp.98-102 (1994)

\bibitem[16]{Clarke95} N. Clarke,
{\it Prot. Sci.} {\bf 4} pp. 2269-2278 (1995)

\bibitem[17]{Thomas96} D. Thomas, G. Casari, C. Sander,
{\it Prot. Eng.} {\bf 9} pp. 941-948 (1996)

\bibitem[18] {Larson2000} S. Larson, A. Di Nardo, A. Davidson,
{\it J. Mol. Biol.} {\bf 303} pp. 433-446 (2000)

\bibitem[19] {Mutual}
Pairwise covariation measures used to date in 
the literature, such as mutual information \cite{Cover91},
are based on the
pairwise probability distribution, $P^{\alpha \beta}_{ij}$, for amino acid
$\alpha$ to occur at sequence position $i$ and amino acid $\beta$ to occur
at sequence position $j$ in an alignment of sequences assumed to
fold to a common structure. Mutual information between
a pair of positions, $MI(ij)$,  can be
defined in terms of $P^{\alpha \beta}_{ij}$ as
$$ MI(ij) =  \sum_{\alpha \beta} P^{\alpha \beta}_{ij} log P^{\alpha \beta}_{ij} /(P^{\alpha}_i P^{\beta}_j) $$
where $P^{\alpha}_i$, respectively $P^{\beta}_j$ are the
single site distributions at positions $i$, respectively $j$, determined
by appropriate summing over the pair distribution.
Other work utilize somewhat different algebraic forms to quantify covariation 
\cite{Lockless99, Larson2000},
but ultimately all published methods manipulate 
probability distributions that involve just two sites at a time,
inherently treating each residue pair of interest
as physically isolated from all other sites in the system.

\bibitem[20]{Cover91} T. Cover,  J. Thomas,
{\it Elements of Information Theory}
Wiley Series in Telecommunications,
John Wiley and Sons (1991)

\bibitem[21] {Lockless99} S. Lockless, R. Ranganathan,
{\it Science} {\bf 286} pp. 295-299 (1999)

\bibitem[22]{InsertBacknote}
A derivation of the maximum entropy distribution for discrete data (i.e. for amino acids),
with application to ``toy'' (simulated) protein models, can be found in \cite{Lapedes97}. Independent work invoking maximum entropy  principles to describe protein
sequences \cite{Saven1} assumed independent sites, and
investigated a different set of issues than those considered here.


\bibitem[23] {Lapedes97}  A. Lapedes,  B. Giraud, L. Liu, G. Stormo,
``Correlated Mutations In Models of Protein Sequences: Phylogenetic and Structural Effects''
in {\it Proceedings of the IMS/AMS 1997 International Conference on Statistics
in Computational Molecular Biology (Seattle 1997)}, published as
Vol 33, Monograph Series of the Inst. for Mathematical Statistics (Hayward Press)

\bibitem[24] {Saven1} H. Kono,  J. Saven,
{\it J. Mol. Biol.} {\bf 306} pp. 607-628(2001)


\bibitem[25]{Backnote4} 
Maximizing the likelihood of the given data (a function of the $\lambda's$)
by gradient ascent yields 
the following iterative procedure for successively
changing the $\lambda's $ by an amount $ \Delta \lambda$ off an initial value 

$$ \Delta \lambda^{\alpha \beta}_{ij} = - \epsilon ({\overline {x^{\alpha}_i x^{\beta}_j}} - <x^{\alpha}_i x^{ \beta}_j> ) $$

$$ \Delta \lambda^{\alpha }_{i} = - \epsilon ({\overline {x^{\alpha}_i }} - <x^{\alpha}_i > ) \, .
$$

Here, ${\overline {x^{\alpha}_i x^{\beta}_j}}$ and ${\overline {x^{\alpha}_i }}$
represent the frequency counts of pairs and single amino 
acids, respectively, in the given sequence data set of interest;
and $<x^{\alpha}_i x^{ \beta}_j>$ and
$<x^{\alpha}_i >$ represent the second and first order moments, respectively,
calculated in the distribution defined by Eqns.(1) and (2).
$\epsilon$ is a small proportionality coefficient, which we set to $0.01$.
The $\lambda's$ are initialized with the $\lambda^{\alpha \beta}_{ij}$
interaction terms set to zero, and the $\lambda^{\alpha}_{i} $ terms
chosen to match the single site probabilities of the given data.
Further algorithmic and numerical details are given in the
supplemental material \cite{SupplementalMaterial}.

\bibitem[26] {Metropolis} N. Metropolis, A. Rosenbluth, M. Rosenbluth, A. Teller,
E. Teller
{\it J. Chem Phys.} {\bf 21} p. 1087 (1953)

\bibitem[27]{BoltzmannMachines} G. Hinton, T. Sejnowski,
``Learning and Relearning in Boltzmann Machines''
in {\it Parallel Distributed Processing}, eds. D. Rumelhart and J. McClelland
MIT Press, Cambridge MA (1986)

\bibitem[28]{GM} S. Lauritzen, {\it Graphical Models}
Oxford Science Publications, Clarendon Press (1996)

\bibitem[29]{Backnote1} Pfam domains that were used follow, in format 
(Pfam name: Pdb designator, Pfam domain size, number of domain positions used, number of sequences)

(Ubiquitin family: 1ubi, domain size=82, length=76, number of sequences=555);
(Homeobox domain: 1fjl, domain size=57, length=57, number of sequences=1777);
(SH3 domain: 1shg, domain size=57, length=55, number of sequences=691);
(EF hand: 3icb, domain size=29, length=29, number of sequences=2159);
(Kunitz/Bovine pancreatic trypsin inhibitor domain: 6pti, domain size=51, length=51, number of sequences=236;)
(Ank repeat: 1awc, domain size=33, length=33, number of sequences=2220;)
(Helix-loop-helix DNA-binding domain: 1mdy, domain size=57, length=52, number of
 sequences=666)
(SH2 domain: 1sha, domain size=79, length=77, number of sequences=567);
(4Fe-4S binding domain: 5fd1, domain size=24, length=24, number of sequences=1209;)
(TPR domain: 1a17, domain size=34, length=34, number of sequences=2507);
(PDZ domain: 1be9, domain size=83, length=81, number of sequences=721).

Sequence positions
defining each folding domain were selected by using the ``match state''
positions of the associated Pfam Hidden Markov Model, available from the Pfam
data base entry for each domain. ``domain size'' is the number of Pfam match
states. Positions in a domain containing indels with less
than fifty percent valid amino acids in the position were deleted from the domain definition.
Positions in a domain containing indels, but with more than fifty percent 
valid amino acids in the position, had the indels ``filled in'' with amino 
acids drawn randomly from the single site amino acid probability distribution
for each such position. For simplicity,
sequence weighting to attempt to address phylogenetic relatedness of
sequences  was not employed. Hennikoff sequence weighting was not found 
\cite{Larson2000} to yield significant improvements, although
conceivably other approaches might be fruitful.

%
%
%

\bibitem[30] {PDB} H. Berman, J. Westbrook, Z. Feng, G. Gilliland, T. Bhat
H. Weissig, I. Shindyalov, P. Bourne,
{\it Nuc. Acids. Res.} {\bf 28} pp. 235-242  (2000)


\bibitem[31]{Parcorr}  N. Draper, H. Smith, {\it Applied Regression Analysis} second edition, Wiley, New York (1981)

\bibitem[32] {Afonnikov97} Linear partial correlation analysis
has been applied to correlations of physico-chemical properties 
of amino acids within secondary structure elements,
D. Afonnikov, Y. Kondrakhin, I. Titov, N. Kilchanov
``Detecting Direct Correlations Between Positions in Multiple Alignments
of Amino Acid Sequences''
in Mewes, H.W. and Firhsman, D. (eds) {\it Computer Science and Biology.
Genome Informatics: Function, Structure, Phylogeny}, Proc. of the German
Conference on Bioinformatics, pp. 87-98 (1997)

\bibitem[33]{Backnote5}
The formula for conditional mutual information between site $i$ and $j$ is
$$ CMI(i,j) = \sum_{x_i x_j X_r} P(x_i x_j X_r) log \frac {P(x_i x_j | X_r)}
{P(x_i|X_r) P(x_j|X_r) } $$
where $x_i$ denotes the residue  at $i$, $x_j$ the residue at $j$,
and $X_r$, of length $L-2$,
denotes the rest of the sequence of amino acids at all sites
other than $i$ and $j$. $P(x_i x_j X_r)$ is merely the joint probability,
Eqns. (1,2), of the complete $L$ long sequence,
while $P(x_i x_j | X_r)$ is the conditional probability of a residue pair
at $i$ and $j$, given the rest of the $L-2$ residues, $X_r$.
Note that these probabilities are functions of the full set 
of $\lambda$ parameters,
i.e., $CMI(i,j)$ in contrast to $MI(i,j)$, involves interactions between 
all residues in the protein. The average of the $log$ expression, above,
can be accurately approximated by Monte Carlo importance sampling \cite{Metropolis}.

\bibitem[34] {Saitoh93} S. Saitoh, T. Nakai, K. Nishikawa,
{\it Proteins} {\bf 15} pp. 191-204 (1993)

\bibitem[35] {Skolnick98a} A. Ortiz, A. Kolinski, J. Skolnick,
{\it J. Mol. Biol}. {\bf 277} ppp. 419-448 (1998)


\bibitem[36]{Backnote6} This is closely related to
specificity-sensitivity plots and the use of ROC curves  for
evaluating predictive power, see e.g.
{\it Science} {\bf 240} pp 1285-93 (1988)

\bibitem[37]{Backnote7} We thank Stefan Larson for provision of
perl scripts that calculate the pairwise covariation measure
in \cite{Larson2000}.

\bibitem[38]{SupplementalMaterial} Supporting online material, to be
available at a website, is presently included in the Appendix
to this manuscript.

\bibitem[39] {Fersht90} A. Horovitz, A. Fersht,
  {\it J. Mol. Biol.} {\bf 214} pp. 613-617 (1990)

\bibitem[40]{Backnote8}
When the $\lambda_{ij}^{\alpha \beta}$ interaction terms are restricted to
be zero, 
the remaining single site terms
may be simply written 
in terms of the single site probabilities of the residues at each position
in the sequence alignment: $\lambda_i^{\alpha}=-log \lbrack P(X_i^\alpha)\rbrack$.
Mutations to residues occurring more frequently in the alignment
would thus be predicted to be stabilizing, an observation that
is often found to be true in experiments, but with frequent exceptions. Allowing
$\lambda_{ij}^{\alpha \beta}$ to be nonzero addresses this 
and other co-operative effects of multiple mutations.

\bibitem[41]{Fersht99} Q. Wang, A. Buckle, N. Foster, C. Johnson, A. Fersht,
   {\it Prot. Sci.} {\bf 8} pp. 2186-2193 (1999)

\bibitem[42]{Baker2000}  B. Kuhlman, D. Baker,
   {\it Proc. Natl. Acad. Sci. USA} {\bf 97} pp. 10383-10388 (2000)


\bibitem[43]{Baker97}  D. Riddle, J. Santiago, S. Bray-Hall, N. Doshi, G. Grantcharova, Q. Yi, D. Baker,
{\it Nature Str. Biol.} {\bf 4} pp. 805-809 (1997)

\bibitem[44]{Szostak2001} A. Keefe, J. Szostak,
   {\it Nature} {\bf 410} pp. 715-718 (2001)




\bigskip 

The research of Lapedes and Jarzynski was supported by the Department of Energy
under contract W-7405-ENG-36. The hospitality of the Santa Fe Institute where 
part of this work was performed is gratefully acknowledged. 
We thank Gary Stormo for very helpful feedback as these ideas evolved, and
Steven Lockless, Rama Ranganathan, Stefan Larson and Tom Woolf for useful
discussions.




\end{thebibliography}
\end{document}